\documentstyle[12pt,aaspp4]{article}
\begin{document}
\def\etal{{\it et al.}}
\def  \vs          {{\it vs.} }
\def  \eg          {{\rm e.g.}}
\def\G{$\Gamma$}
\def\egs{{\it EGRET}~}
\def\gr{$\gamma$-ray}
\def\IUE{{\it IUE}}
%
%
%
\def  \La          {\ifmmode {\rm Ly}\alpha \else Ly$\alpha$\fi}
\def  \Ka          {\ifmmode {\rm K}\alpha \else K$\alpha$\fi}
\def  \Lb          {\ifmmode {\rm L}\beta \else L$\beta$\fi}
\def  \Ha          {\ifmmode {\rm H}\alpha \else H$\alpha$\fi}
\def  \Hb          {\ifmmode {\rm H}\beta \else H$\beta$\fi}
\def  \Pa          {\ifmmode {\rm P}\alpha \else P$\alpha$\fi}
\def  \HeI        {\ifmmode {\rm He}\,{\sc i}\,\lambda5876
                     \else He\,{\sc i}\,$\lambda5876$\fi}
\def  \HeII        {\ifmmode {\rm He}\,{\sc ii}\,\lambda4686
                     \else He\,{\sc ii}\,$\lambda4686$\fi}
\def  \HeIIa        {\ifmmode {\rm He}\,{\sc ii}\,\lambda1640
                     \else He\,{\sc ii}\,$\lambda1640$\fi}
\def  \HeIIb        {\ifmmode {\rm He}\,{\sc ii}\,\lambda1085
                     \else He\,{\sc ii}\,$\lambda1085$\fi}
\def  \CII        {\ifmmode {\rm C}\,{\sc ii}\,\lambda1335
                     \else C\,{\sc ii}\,$\lambda1335$\fi}
\def  \CIII        {\ifmmode {\rm C}\,{\sc iii}\,\lambda977
                     \else C\,{\sc iii}\,$\lambda977$\fi}
\def  \CIIIb       {\ifmmode {\rm C}\,{\sc iii]}\,\lambda1909
                     \else C\,{\sc iii]}\,$\lambda1909$\fi}
\def  \CIV         {\ifmmode {\rm C}\,{\sc iv}\,\lambda1549
                     \else C\,{\sc iv}\,$\lambda1549$\fi}
\def  \bOIIIb       {\ifmmode {\rm [O}\,{\sc iii]}\,\lambda5007
                     \else [O\,{\sc iii]}\,$\lambda5007$\fi}
\def  \OIIIb       {\ifmmode {\rm O}\,{\sc iii]}\,\lambda1663
                     \else O\,{\sc iii]}\,$\lambda1663$\fi}
\def  \OVIII        {\ifmmode {\rm O}\,{\sc viii}\,653~{\rm eV}
                     \else O\,{\sc viii}\,$653~{\rm eV}$\fi}
\def  \OVII        {\ifmmode {\rm O}\,{\sc vii}\,568~eV
                     \else O\,{\sc vii}\,$568~{\rm eV}$\fi}
\def  \OVI         {\ifmmode {\rm O}\,{\sc vi}\,\lambda1035
                     \else O\,{\sc vi}\,$\lambda1035$\fi}
\def  \OIVb         {\ifmmode {\rm O}\,{\sc iv]}\,\lambda1402
                     \else O\,{\sc iv]}\,$\lambda1402$\fi}
\def  \bOIIb       {\ifmmode {\rm [O}\,{\sc ii]}\,\lambda3727
                     \else [O\,{\sc ii]}\,$\lambda3727$\fi}
\def  \bOIb       {\ifmmode {\rm [O}\,{\sc i]}\,\lambda6300
                     \else [O\,{\sc i]}\,$\lambda6300$\fi}
\def  \NII         {\ifmmode {\rm N}\,{\sc ii}\,\lambda1084
                     \else N\,{\sc ii}\,$\lambda1084$\fi}
\def  \NIII         {\ifmmode {\rm N}\,{\sc iii}\,\lambda990
                     \else N\,{\sc iii}\,$\lambda990$\fi}
\def  \NIIIb         {\ifmmode {\rm N}\,{\sc iii]}\,\lambda1750
                     \else N\,{\sc iii]}\,$\lambda1750$\fi}
\def  \NIVb         {\ifmmode {\rm N}\,{\sc iv]}\,\lambda1486
                     \else N\,{\sc iv]}\,$\lambda1486$\fi}
\def  \NV          {\ifmmode {\rm N}\,{\sc v}\,\lambda1240
                     \else N\,{\sc v}\,$\lambda1240$\fi}
\def  \MgII        {\ifmmode {\rm Mg}\,{\sc ii}\,\lambda2798
                       \else Mg\,{\sc ii}\,$\lambda2798$\fi}
\def  \CVI        {\ifmmode {\rm C}\,{\sc vi}\,368~eV
                       \else C\,{\sc vi}\,368~eV\fi}
\def  \SiIV         {\ifmmode {\rm Si}\,{\sc iv}\,\lambda1397
                     \else Si\,{\sc iv}\,$\lambda1397$\fi}
\def  \bFeXb       {\ifmmode {\rm [Fe}\,{\sc x]}\,\lambda6734
                       \else [Fe\,{\sc x]}\,$\lambda6734$\fi}
\def  \MgX        {\ifmmode {\rm Mg}\,{\sc x}\,\lambda615
                       \else Mg\,{\sc x}\,$\lambda615$\fi}
\def  \MgXI        {\ifmmode {\rm Mg}\,{\sc xi}\,1.34~keV
                       \else Mg\,{\sc xi}\,1.34~keV\fi}
\def  \MgXII      {\ifmmode {\rm Mg}\,{\sc xii}\,1.47~keV
                     \else Mg\,{\sc xii}\,1.47~keV\fi}
\def  \bNeVb      {\ifmmode {\rm [Ne}\,{\sc v]}\,\lambda3426
                     \else [Ne\,{\sc v]}\,$\lambda3426$\fi}
\def  \NeVIII      {\ifmmode {\rm Ne}\,{\sc viii}\,\lambda774
                     \else Ne\,{\sc viii}\,$\lambda774$\fi}
\def  \SiVIIa      {\ifmmode {\rm Si}\,{\sc vii}\,\lambda70
                     \else Si\,{\sc vii}\,$\lambda70$\fi}
\def  \NeVIIa      {\ifmmode {\rm Ne}\,{\sc vii}\,\lambda88
                     \else Ne\,{\sc vii}\,$\lambda88$\fi}
\def  \NeVIIIa      {\ifmmode {\rm Ne}\,{\sc viii}\,\lambda88
                     \else Ne\,{\sc viii}\,$\lambda88$\fi}
\def  \NeIX      {\ifmmode {\rm Ne}\,{\sc ix}\,915~eV
                     \else Ne\,{\sc ix}\,915~eV\fi}
\def  \NeX      {\ifmmode {\rm Ne}\,{\sc x}\,1.02~keV
                     \else Ne\,{\sc x}\,1.02~keV\fi}
\def  \SiXII        {\ifmmode {\rm Si}\,{\sc xii}\,\lambda506
                       \else Si\,{\sc xii}\,$\lambda506$\fi}
\def  \SiXIII      {\ifmmode {\rm Si}\,{\sc xiii}\,1.85~keV
                     \else Si\,{\sc xiii}\,1.85~keV\fi}
\def  \SiXIV      {\ifmmode {\rm Si}\,{\sc xiv}\,2.0~keV
                     \else Si\,{\sc xiv}\,2.0~keV\fi}
\def  \SXV      {\ifmmode {\rm S}\,{\sc xv}\,2.45~keV
                     \else S\,{\sc xv}\,2.45~keV\fi}
\def  \SXVI      {\ifmmode {\rm S}\,{\sc xvi}\,2.62~keV
                     \else S\,{\sc xvi}\,2.62~keV\fi}
\def  \ArXVII      {\ifmmode {\rm Ar}\,{\sc xvii}\,3.10~keV
                     \else Ar\,{\sc xvii}\,3.10~keV\fi}
\def  \ArXVIII      {\ifmmode {\rm Ar}\,{\sc xviii}\,3.30~keV
                     \else Ar\,{\sc xviii}\,3.30~keV\fi}
\def  \FeI_XVI      {\ifmmode {\rm Fe}\,{\sc 1-16}\,6.4~keV
                     \else Fe\,{\sc 1-16}\,6.4~keV\fi}
\def  \FeXVII_XXIII     {\ifmmode {\rm Fe}\,{\sc 17-23}\,6.5~keV
                     \else Fe\,{\sc 17-23}\,6.5~keV\fi}
\def  \FeXXV      {\ifmmode {\rm Fe}\,{\sc xxv}\,6.7~keV
                     \else Fe\,{\sc xxv}\,6.7~keV\fi}
\def  \FeXXVI      {\ifmmode {\rm Fe}\,{\sc xxvi}\,6.96~keV
                     \else Fe\,{\sc xxvi}\,6.96~keV\fi}
\def  \FeLa     {\ifmmode {\rm Fe}\,{\sc L}\,0.7-0.8~keV
                     \else Fe\,{\sc L}\,0.7-0.8~keV\fi}
\def  \FeLb     {\ifmmode {\rm Fe}\,{\sc L}\,1.03-1.15~keV
                     \else Fe\,{\sc L}\,1.03-1.15~keV\fi}
%
%
\def  \L46         {$ L_{46} $}
\def  \Mo          {$ M_{\odot} $}     
\def  \Lo          {$ L_{\odot} $}     
\def  \Fn          {$ F_{\nu} $}
\def  \Ln          {$ L_{\nu} $}
\def  \LEdd        {$ L_{\rm Edd} $}    
\def  \MBH         {$ M_{\rm BH} $}     
\def  \m9          {$ m_9 $}
\def  \mdot        {$ \dot{m} $}        
\def  \Mdot        {$ \dot{M} $}        
\def  \Te          {$ T_{\rm e} $}      
\def  \Tex         {$ T_{\rm ex} $}     
\def  \Teff        {$ T_{\rm eff} $}    
\def  \THIM        {$ T_{\rm HIM} $}    
\def  \TC          {$ T_{\rm C} $}      
\def  \Ne          {$ N_{\rm e} $}      
\def  \nh          {$ n_{\rm H} $}      
\def  \N10         {$ N_{10} $}
\def  \Ncol        {$ N_{\rm col} $}    
\def  \rin         {$ r_{\rm in} $}
\def  \rout        {$ r_{\rm out} $}
\def  \rav         {$ r_{\rm av} $}
\def  \jc          {$ j_c(r) $}
\def  \Rc          {$ R_c(r) $}
\def  \Ac          {$ A_c(r) $}
\def  \nc          {$ n_c(r) $}
\def  \Cr          {$ C(r) $}
\def  \Cf          {$ \rm C_f $}     
\def  \El          {$ E_l(r) $}
\def  \el          {$ \epsilon_l(r) $}
\def  \PSI         {$ \Psi (t) $}
\def  \me          {$ m_{\rm e} $}            
\def  \mp          {$ m_{\rm p} $}            
\def  \rBLR        {$ r_{\rm BLR} $}          
\def  \rNLR        {$ r_{\rm NLR} $}          
\def  \aox         {$ \alpha_{ox} $}          %
\def  \Ux          {$ \rm U_{\rm x} $}  
%
%
\def  \kms         {\hbox{km s$^{-1}$}}          
\def  \ergs        {\hbox{erg s$^{-1}$}}              
\def  \ergsHz      {\hbox{erg s$^{-1}$ Hz$^{-1}$}}   
\def  \cc          {\hbox{cm$^{-3}$}}
\def  \cmii        {\hbox{cm$^{-2}$}}
\def  \cms         {\hbox{cm s$^{-1}$}}      
\def  \mic         {$\mu$m}
\def  \vs          {{\it vs.} }
\def  \etal        {{\it et al.}}
%
%
\newbox\grsign \setbox\grsign=\hbox{$>$} \newdimen\grdimen \grdimen=\ht\grsign
\newbox\simlessbox \newbox\simgreatbox \newbox\simpropbox
\setbox\simgreatbox=\hbox{\raise.5ex\hbox{$>$}\llap
     {\lower.5ex\hbox{$\sim$}}}\ht1=\grdimen\dp1=0pt
\setbox\simlessbox=\hbox{\raise.5ex\hbox{$<$}\llap
     {\lower.5ex\hbox{$\sim$}}}\ht2=\grdimen\dp2=0pt
\setbox\simpropbox=\hbox{\raise.5ex\hbox{$\propto$}\llap
     {\lower.5ex\hbox{$\sim$}}}\ht2=\grdimen\dp2=0pt
\def\simgreat{\mathrel{\copy\simgreatbox}}
\def\simless{\mathrel{\copy\simlessbox}}
\def\simprop{\mathrel{\copy\simpropbox}}
\def \GINGA  {{\it GINGA}}
\def \Ginga  {{\it Ginga}}
\def \ASCA   {{\it ASCA}}
\def \BBXRT  {{\it BBXRT}}
\def \ROSAT  {{\it ROSAT}}
\def \XTE    {{\it XTE}}
\def \XLi    {XL$_1$}
\def \XLii   {XL$_2$}
\def \XLiii  {XL$_3$}  
\def \XMi    {XM$_1$}
\def \XMii   {XM$_2$}
\def \XMiii  {XM$_3$}
%
%
%
\def \aap     { {\it Astr. Ap.}, }
\def \aaps    { {\it Astr. Ap. Suppl.}, }
\def \aj      { {\it A. J.}, }
\def \apj      { {\it Ap. J.}, }
\def \apjl     { {\it Ap. J. (Letters)}, }
\def \apjs    { {\it Ap. J. Suppl.}, }
\def \iau      { {\it International Astronomical Union} }
\def \mnras   { {\it M.N.R.A.S.}, }
\def \pasp    { {\it Pub.A.S.P.}, }
\def \BAAS  { {\it Bulletin of the American Astronomical Society}}
\def \Icarus  { {\it Icarus}}
\def \JMolSpec  { {\it J. Molec. Spectrosc.}}
\def \JQSRT  { {\it J. Quant. Spectrosc. Rad. Trans.}}
\def \Science  { {\it Science}}
\def \JGR  { {\it J. Geophys. Res.}}
\def \PhDthesis  { {\it Ph. D. thesis}}
\def \JOptSocAm  { {\it J. Opt. Soc. Am.}}
%
\title{Ultraviolet Lines and Gas Composition in NGC 1068}
%
%
\author{Hagai Netzer \altaffilmark{1}}
\altaffiltext{1}{School of Physics and Astronomy and the Wise
Observatory, The Raymond and  Beverly Sackler Faculty of Exact Sciences,
Tel-Aviv University, Tel-Aviv 69978, Israel.}
%
%
%

\begin{abstract}
Recent ultraviolet and X-ray observations 
 suggest that the Fe/O ratio in the NLR gas in
NGC~1068 is abnormally high.
The X-ray analysis, which is presented elsewhere, suggests a large
Fe/H and the data shown and discussed here, in particular the extremely  weak
 \OIIIb\ line,  argue for small O/N and
O/C. Models to support this claim are shown and discussed.
They include improved reddening estimates, dusty and  dust-free calculations, and
a range of abundances.
 The   anomelous composition makes NGC 1068 unique
among Seyfert galaxies and an unusual laboratory for investigating  metal
enrichment and depletion. 
\end{abstract}

\section{Introduction}
NGC 1068 is perhaps the ideal benchmark for studying 
gas composition in active galaxies. Its proximity, high luminosity
and inclination make spectroscopic studies easier than in most  known 
Seyferts. As a result, the available spectroscopic information is of unusual
quality and coverage. The optical and UV spectrum have been studied in numerous
papers (Snijderes, Netzer \& Boksenber; 1986, and references therein)
 and the extreme UV spectrum has been observed
by HUT (Kriss \etal\ 1992 and this volume). Dozens of emission lines have been
measured and soon to be published papers will include the first  attempt at  
high spatial resolution long-slit spectroscopy. This  will provide 
line ratios  in  regions of different levels of ionization and will enable
a  meaningful comparison with multi-component models. 

The  X-ray information on this source is even more remarkable.
 BBXRT (Marshall
\etal\ 1993) and ASCA (Ueno \etal\ 1994)
observations  revealed the presence of several strong X-ray lines.
The iron \Ka\ complex is clearly split into three components (Marshall \etal\
1993), suggesting the presence of highly ionized X-ray heated gas. There is also
evidence for several soft X-ray lines with equivalent width  of at least
50--100 eV. This is the first and  best evidence for soft X-ray lines in any 
AGN.
The anlysis of the X-ray  
 lines can be used
to put limits on the Fe/O abundance ratio. The complimentary information
from  UV studies    provides limits on  O/N and O/C. 

This paper discusses the gas composition in NGC 1068.
 The X-ray aspects are
 reviewed  in \S2. A detailed  analysis of the UV spectrum is given
in \S3 with emphasis on recent HST and HUT observations. \S4 summarizes the most
important findings of this work.

\section{The X-ray spectrum of NGC 1068}
The most detailed X-ray investigations of NGC~1068 are by Marshall
\etal\ (1993) and Ueno \etal\ (1994). Previous Ginga observations are
described by Awaki \etal\ (1990)
 and ROSAT imaging is described in Wilson \etal\ (1992). 
The analysis of the BBXRT data, by Marshall \etal\ (1993), suggests 
two different components that contribute to the iron \Ka\ complex.
1. A
``warm'' component, with a temperature of about $2 \times 10^5$K, which produces
the 6.4 keV line. This  gas is photoionized by the nuclear source and is also
the  ``mirror'' reflecting the optical-UV continuum and the broad 
Balmer lines. 2. A hot,  ${\rm T} \simeq 4 \times 10^6$K component,   
producing the  \FeXXV\ and
\FeXXVI\ lines. According to Marshall \etal, the two contributes about equally to
the scattered 6.4 continuum. A different combination of components is suggested by
Iwasawa \etal\ (1997). In their scheme, the 6.4 keV line is due to a
Compton thick gas, with very large column density, presumably the thick walls of
the central torus. Iwasawa \etal\  suggested also a second, hot component to explain
the highly ionized iron lines. In their model, the scattered 6.4 keV continuum is mostly due to the Compton-thick gas and there is no explanation of the ``mirror''.

Marshall \etal\ (1993) reported an upper limit on the  \OVIII\  line which indicates
 a large oxygen under abundance. The most significant comparison
is with  iron since the observed  \Ka\ lines enables a simple
determination of Fe/H. The analysis suggests an iron over abundance,
relative to solar,
by a factor of 2--3 and oxygen under abundance by a factor of 3--5. Ueno \etal\
(1994)  confirmed the upper limit on the \OVIII\ line  and the 
BBXRT  measurements of the \Ka\ lines.  They assumed that
most of the soft X-ray flux is from hot, collisionally excited
plasma, in which all metals are  highly depleted.
Netzer and Turner (1997) reanalyzed the ASCA data.  They measured  several
soft X-ray lines, such as \NeIX, \NeX, H-like and helium like lines of magnesium,
silicon, sulphur and argon, and several Fe-L lines, and confirmed the two-component model.  
Marshall \etal\ (1993) have also examined   the UV spectrum of NGC~1068 and
 suggested  more evidence for the small O/H.

\section{The Ultraviolet Spectrum and the O/N abundance}

The observed optical -UV spectrum of NGC~1068 is described in 
 Snijders \etal\ (1986).
 More
recent, shorter wavelength (900--1800\AA) observations by HUT are reported in
Kriss \etal\ (1992; see also G. Kriss contribution in this volume). All 
large aperture data  represent the spectrum over a region of 
roughly 20 arc-sec in diameter. The high resolution HST imaging, such as the
one presented in this meeting (e.g.Z. Tsvetanov contribution), clearly emphasize the
limitation of large aperture observations. 

The HST archive contains some high spatial resolution spectroscopy
of this galaxy. We have used one such unpublished data set, obtained by centering
on an off-nucleus cloud, to
try and separate the high and the low excitation lines. Our line list (Table 1)
is an average of several sources designed to represent an average ionization
cloud. 

\subsection{Dust and Reddening}
The  spectrum of NGC~1068 indicates
a substantial amout of reddening. Snijders \etal\ (1986) used a galactic
type extinction and estimated E$_{\rm B-V}=0.4$ mag., mostly 
from  the \HeII/\HeIIa\ line ratio. The more recent  HUT observations of Kriss \etal\ 
(1992) discovered two helium lines, 
 \HeIIa\ and  \HeIIb. Their ratio, relative to \HeII, is inconsistent
with a single extinction law.
 This has been discussed, extensively, by Kriss
\etal\ (1992) and Ferguson \etal\ (1995). Ferguson \etal\ (1995) suggested a 
double reddening correction approach in
 which all lines with wavelength longer than 1640\AA\
were corrected with one extinction curve and all lines with shorter wavelength
were corrected with a different curve. Unfortunately, there was a mistake in 
applying  this procedure to the data (Ferguson, private communication) and the
reddening corrected line intensities in their table 1 are erroneous.

The difficulty encountered by Kriss \etal\ (1992) and Ferguson \etal\ (1995)
is due to the inconsistency in the observed flux of three HeII lines
 at 4686, 1640 and 1084\AA.
The line intensities are predicted by simple recombination
theory and are very reliable reddening indicators.
Ferguson \etal\ noted that continuum fluorescence (the absorption of continuum
photons by resonance lines) can contribute significantly to several  of the
observed UV transitions. They have  shown that the 
\CIII\ and \NIII\ intensities can considerably increase due to this process. Our
new calculations support this claim. They also show that \NII\ 
is significantly affected by this process. When including continuum 
fluorescence in the calculations of this line,
 with velocity broadening of 500 \kms, we
find that this  can explain up to 50\% of the observed 1084\AA\ flux. This
cures the reddening anomaly and enables a single, self consistent
reddening solution with a galactic-type extinction. We further note that some
other lines can significantly be affected by continuum fluorescence. In particular, in large
column density situations, there can be 
a large contribution at 1304\AA, 1039\AA, 1026\AA, 990\AA, 976\AA, and 972\AA,
from OI resonance lines. Some of the observed flux at around 1035\AA, 977\AA\ and 990\AA,  
can be due to this process, although the absence 
of a strong 1304\AA\ is problematic. Examples are shown in Table 1.

Table 1 lists dereddened line ratios assuming  galactic-type extinction and
E$_{\rm B-V}=35$ mag. 
The calculated  contributions
to the 1084\AA\ feature, suggests this to be the most reliable reddening estimate.
The listed uncertainties
 on the dereddened line intensities 
  represent the 2$\sigma$ intervals, as given in the
original papers,  without taking into 
account the uncertainty in  reddening. The listed intensity of
\La\ has no assigned uncertainty due to the blending under geocoronal \La.  
\begin{table}
\caption{Observed and calculated line intensities}
\begin{minipage}{13cm}
\begin{tabular}{lccc}
\hline
\hline
Line & Dusty NLR\footnote{Galactic dust-to-gas ratio.
            Ionization parameter of
             9.4$\times 10^{-3}$ and ${\rm n_H}$=10$^4$\,\cc.}
             & Dust-free NLR&Dereddened line
              ratios\footnote{E$_{\rm B-V}=0.35$.
              The Observed \Hb\ flux is 1.6$\times 10^{-12}$
             ergs s$^{-1}$ cm$^{-2}$ }\\
\hline
\Ha         &3.3&  2.75  &  2.7-3.0 \\
\Hb         & 1.0&1.0  & 1.0 \\
\La         & 17 & 29.5& 35(?) \\
\HeII       & 0.42 & 0.30& 0.46 - 0.72 \\
\HeIIa       &1.68  &1.9& 4.1 - 4.8 \\
\HeIIb  & 0.25  & 0.29&1.2 - 2.1
              \footnote{blended with \NII, see text} \\ 
\CII        &  0.18 &0.61   & 0.7 - 1.1  \\
\CIII       & 0.77 & 1.5 &  3.4 - 4.7  \\
\CIIIb      & 5.3 & 6.6&4.9 - 6.0  \\
\CIV        & 6.8& 9.8& 7.2 -  8.3 \\ 
\NII        & 0.13& 0.39 & 1.2 - 2.1 \\ 
\NIII       & 0.43 &  0.80& 1.5 - 2.5 \\
\NIIIb      & 0.93 & 0.48& 0.7 - 1.4 \\
\NIVb       & 1.22  &0.53& 0.8 - 1.4  \\
\NV         & 1.1 & 0.63 & 8.4 - 10.1 \\
\bOIIb      &0.24  & 0.12 &0.5 - 0.8 \\
\bOIIIb     & 12.0 & 9.8 &16 - 18   \\
\OIIIb      & 0.46 & 0.30 &$<$0.29 \\
\OIVb+\SiIV & 1.29  &  0.84 & 1.6 - 2.2 \\
\OVI        & 0.22   & 0.19 & 16 - 20 \\
${\rm O}${\sc i}$\lambda1304$ & 0.09& 0.26& \\
${\rm O}${\sc i}$\lambda1026$ & 0.03& 0.20& \\
${\rm O}${\sc i}$\lambda1039$ &0.02 &0.13 & \\
${\rm O}${\sc i}$\lambda989$ & 0.04&0.31 & \\
${\rm O}${\sc i}$\lambda976+972$ &0.01 &0.26 & \\
\MgII       & 1.8  & 3.7 &0.7 - 1.1     \\ 
\hline
\end{tabular}
\end{minipage}
\end{table}

\subsection{Photoionization Calculations}
We have used the photoionization code ION (Netzer 1996 and references therein) to
carry out extensive photoionization modeling of the NLR gas in NGC~1068. We have
assumed various densities and column densities, different geometries and dust
distribution 
 and allowed for abundance variations. The line
list presented in Table 1 is used for comparison and we have tried splitting it
into high and low excitation lines to test the quality of the fit. Two single component
models are shown in the table. They have similar ionization parameter, density
and column density, and differ by the assume amount of internal dust. The
assumed turbulent velocity is 500 \kms.

The continuum used for the calculation is similar to the one used 
 by Marshall \etal\ (1993), except for the soft X-ray part. It
is made of a broken power-law with typical optical-UV slope and
\aox=1.35. This  fits best with the overall energy budget as well
as the observed hard X-ray flux. The main uncertainty is due to the shape of
the unobserved UV bump. As argued by
Wilson \etal\ (1992), 
 there is a likely strong soft-X-ray contribution
from an extended source, presumably a starburst region.
 This
 implies a weaker  soft  X-ray central source, compared with the one used in  Marshall \etal.
 We have considered this possibility and 
have  assumed a hard (F$_{\nu} \propto \nu^{-0.5}$) X-ray continuum
extending from 1 to 50 keV,  normalized to the observed flux at 6.4 keV.
The 1 keV point was connected to the lower energy continuum used by Marshall
\etal.

In anticipation of the unusual composition, and in accord with the X-ray analysis,
we have modeled the NLR gas with He/H=0.1 and
the following dust-free composition:\\
H:C:N:O:Ne:Mg:Si:S:Ar:Fe=
 $10^{4}$:3:1:1.5:1:0.33:0.33:0.16:0.07:1.2.\\ 
For the dusty NLR we have assumed gas phase composition of\\
 H:C:N:O:Ne:Mg:Si:S:Ar:Fe=
   $10^{4}$:1.6:1:1.5:0.6:0.16:0.016:0.054:0.016:0.016. 

Imaging of  NGC~1068, through various narrow band
 filters, clearly show ionization stratification.
 Thus  single component models, like the ones shown in the 
table, with ``mean'' properties, are  poor
representations of the more complex physical 
situation. The emphasis in this work is on the gas composition, and
we do not list most of the calculated models.
 We also point out   that the derived E$_{\rm B-V}$
 is more appropriate to the high ionization component, since it is
based on HeII line ratios. Any composite model 
must take into account the possibility
that the high and the low ionization lines are affected in a different way by
internal dust. In particular,  internal dust 
 is more important in
 high ionization parameter environment (e.g. Laor and Netzer, 1993). For
example, 
in  dusty models with much larger ionization parameter,
 the HeII line ratio is affected too. However, the derived
O/N and O/C abundance ratios (see below)
  are hardly affected.
 
\subsection{Abundance determination}
A key issue in the present study is the abundance of iron and oxygen. 
The iron abundance can be obtained from  X-ray modeling since the 
equivalent width of
the strong \Ka\ lines is a direct measure of Fe/H.
 The oxygen abundance can be determined by comparing oxygen, nitrogen and
carbon emission lines. As argued by Shields (1976), 
several ultraviolet semi-forbidden lines can be used in this analysis because
of their similar excitation and level of ionization. The line
ratios most suitable for our purpose are \OIIIb/\NIIIb\ and \OIIIb/\CIIIb.

The ionization structure of a gas cloud exposed to a typical AGN ionizing 
continuum shows a large overlap of the C$^{+2}$, N$^{+2}$ and  O$^{+2}$ ionization regions. This is illustrated in Fig. 1  for the assumed 
NGC~1068 continuum.
\begin{figure}
\epsfxsize\hsize
\epsfbox{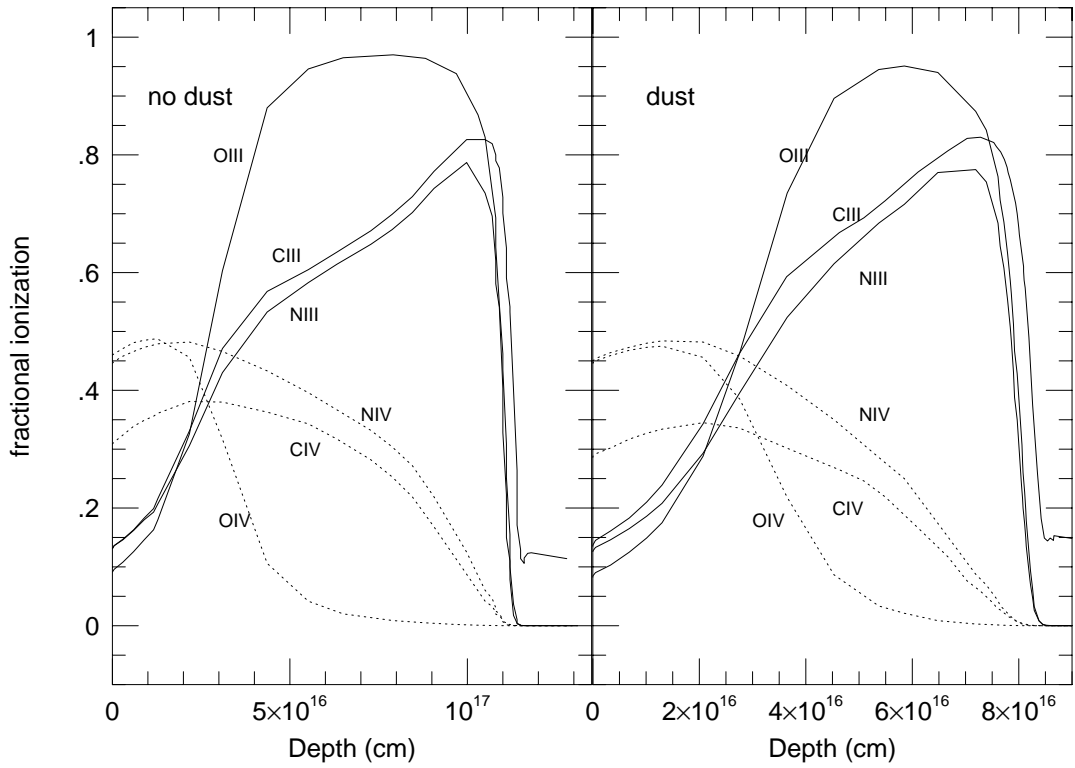}
\caption{Ionization structure for dusty and dust-free models calculated
 for NGC~1068.  Note the almost
    perfect overlap of the  C$^{+2}$, N$^{+2}$ and  O$^{+2}$ 
    zones.}
\end{figure}
Given the similar ionization structure, we can use the observed line ratios,
to deduce  relative abundances. The theoretical ratios are 
\begin{equation}
\frac {{ \rm I(O iii]}\lambda1663  )} { {\rm I(N iii]}\lambda1750  )} =
     0.41 {\rm T_4^{-0.04}} \exp (-0.43/{\rm T_4}) 
       \frac { {\rm N(O^{+2})} } { {\rm N(N^{+2}) } }
\end{equation}
and 
\begin{equation}
\frac {{ \rm I(O iii]}\lambda1663  )} { {\rm I(C iii]}\lambda1909  )} =
     0.15 {\rm T_4^{0.06}} \exp (-1.11/{\rm T_4}) 
       \frac { {\rm N(O^{+2})} } { {\rm N(C^{+2}) } } \,\, ,
\end{equation}
where ${\rm T_4}$ is the  temperature in units of 10$^4$K. 
We  further assume (Fig. 1) that 
N(O)/N(N)$\simeq {\rm N(O^{+2})} / {\rm N(N^{+2}) }$
and
N(O)/N(C)$\simeq {\rm N(O^{+2})} / {\rm N(C^{+2}) }$.
The observed intensities (Table 1), combined with a typical ${\rm T_4}$=1.5,
gives  
N(O)/N(N)=$<1.0$ and 
and N(O)/N(C)=$<0.73$. These ratios are much below the solar values and confirm
the large oxygen under abundance in this source. Metal depletion
on grains cannot provide an explanation to this ratio unless the gas and dust
 composition are vastly different from their  ISM values. 

\section{Discussion and Conclusions}

Analysis of the composite spectrum of NGC~1068 clearly demonstrates the
unusual oxygen line ratios. The examples shown in this work suggest that
the NLR gas has extremely low  oxygen abundance. The models listed in Table 1 are not intended to 
represent the full physical conditions, only to demostrate the more important
processes. They confirm the need for large turbulent motions (Ferguson \etal;
1995) and eliminate
much of the problem raised by Kriss \etal\ regarding the intensity of
\CIII\ and \NIII. They fall short of explaining the strong \OVI\ and 
\NV\ lines and suggest that 
higher ionization components, perhaps of optically thin gas, may be important.
In summary, future models   of NGC~1068
must take into account the following:
\begin{enumerate}
\item
The ionized gas is stratified and regions of different excitation must be combined
in order to mimic the observed large aperture spectrum.
\item
Large turbulent motions
 make continuum fluorescence an important process and can significantly
 increases the
intensity of various resonance lines. This  helps to explain the strong \CIII\ and 
\NIII\ lines (Ferguson \etal; 1995), the anomalous 1084\AA\ feature (this work)
and perhaps the strength of some ultraviolet OI lines (this work). The process
is likely to be more important in Seyfert 2s, where the central continuum is 
obscured from direct view.
\item
When taking into account the \NII\ line, there is a good agreement with
 a single galactic-type
extinction law, with E$_{\rm B-V} \simeq 0.35$ mag.
\item 
Dust mixed with the NLR gas
is likely to be important in the NLR.
\item
The absence of the \OIIIb\ line   suggests 
oxygen under abundance, compared with nitrogen and carbon. This is in line
with  indications from  X-ray (Netzer and Turner; 1997) and millimeter 
(Sternberg, Genzel \& Tacconi; 1994) observations. It is interesting
to note the  intensity of this line 
 in other Seyfert galaxies.
  \OIIIb\ is clearly observed in both NGC~4151 (Ulrich \etal\
1985) and NGC~5548 (Crenshaw \etal\ 1993) during times 
when the broad emission lines are weak. The line
is also observed in the Seyfert 2 galaxy NGC~5525 (Tsvetanov \etal, in preparation)
where it is stronger than \NIIIb. In all these
 cases, there are indications for solar O/C and O/N. Finally, our recent HST 
study of the Seyfert 2 galaxy  NGC~3393 (Cooke \etal; 1997)
 does not show the \OIIIb\ line. However, the FOS spectrum is of moderate quality
and the observed intensity is consistent with solar O/C and O/N.
\end{enumerate}

{\it Acknowledgements:} This research was supported in part by a grant from the
Israel Science Foundation.

\end{document}